\PassOptionsToPackage{english}{babel}
\documentclass[10pt,twocolumn,prl,showpacs,preprintnumbers,superscriptaddress,amsmath,amssymb]{revtex4} 
\usepackage{graphicx, hyperref, color, bm} 
\usepackage{enumitem}
\newcommand{\BE}{\begin{equation}}
\newcommand{\EE}{\end{equation}}
\def\BEA{\begin{eqnarray}}
\def\EEA{\end{eqnarray}}
\def\oa{{O_{a}}}
\def\ob{{O_{b}}}
\def\oc{{O_{c}}}

\def\on{{O_{n}}}
\def\oe{{O_{\epsilon}}}

\def\cpla{{\lambda_{a}}}
\def\cplb{{\lambda_{b}}}
\def\cplc{{\lambda_{c}}}

\def\cpln{{\lambda_{n}}}
\def\cple{{\lambda_{\epsilon}}}

\def\const{{\Omega}}

\def\jrng{K} 
\def\jdga{J_{d}}
\def\jdgb{J_{d}'}
\def\jdgl{\tilde{J}_{d}}
\def\jnnn{J_2}
\def\jvrt{J_{\perp}}
\begin{document}
\title{Competing phases in spin ladders with ring exchange and frustration}
\author{Alexandros Metavitsiadis}
\affiliation{Department of Physics and OPTIMAS Research Center, University of Kaiserslautern,
D-67663 Kaiserslautern, Germany}
\affiliation{Institute for Theoretical Physics
Technical University Braunschweig, 
D-38106 Braunschweig, Germany}
\author{Sebastian Eggert} 
\affiliation{Department of Physics and OPTIMAS Research Center, University of Kaiserslautern,
D-67663 Kaiserslautern, Germany}
\date{\today}
\begin{abstract}
The ground state properties of  spin-1/2 ladders are studied, 
emphasizing the role of frustration and ring exchange coupling. 
We present a unified field theory for ladders with general  
coupling constants and geometry.   Rich phase diagrams can be deduced
by using a renormalization group calculation for ladders with 
in--chain next nearest neighbor interactions and plaquette ring 
exchange coupling.   In addition to established phases such as Haldane, rung singlet, 
and dimerized phases, we also observe a surprising instability towards an incommensurate
phase for weak interchain couplings, which 
is characterized by an exotic coexistence of self-consistent ferromagnetic 
and anti-ferromagnetic order parameters.

\end{abstract}
%
%
%
\maketitle
%
%
%
%
%
%
Spin ladders are heavily studied prototypical models 
exemplifying the role of quantum fluctuations 
in low dimensional quantum systems \cite{rice-low-dimensional-magnets, 
Dagotto02021996, 978-3-540-21422-9,  
PhysRevB.53.52,PhysRevB.53.8521, JPSJ.64.1967, JPSJ.72.1022, 
PhysRevB.66.134423, PhysRevB.69.094431, PhysRevB.88.075132,
PhysRevB.88.104403,
PhysRevB.62.14965,PhysRevB.61.8871,PhysRevLett.93.127202,
PhysRevB.81.064432,PhysRevB.86.075133,PhysRevB.73.224433,PhysRevB.77.205121,
PhysRevB.59.7973, PhysRevB.72.104435,PhysRevB.73.214427,PhysRevB.84.144407,PhysRevB.89.241104,
PhysRevB.77.214418, 1367-2630-14-6-063019}.
Already in their simplest form they show some of the most discussed 
quantum many-body properties, as for example a spin liquid or a Haldane gapped state with 
topological string order parameter. 
Theoretical studies have identified 
a number of remarkable ground state phases, but  
given the large variety of possible tuning parameters it is likely that 
this list is far from
complete.  At the same time experimental progress on novel materials 
\cite{PhysRevLett.79.151, PhysRevLett.111.107202, PhysRevLett.110.187201} as
well as advancements in the field of optical lattices \cite{PhysRevB.88.165101,
PhysRevA.87.043625, nature09994, PhysRevLett.110.184102} give renewed interest
in spin ladders in their own right beyond the 
larger
effort to gain a better insight into complex phases of frustrated two-dimensional (2D) models.
This work now focuses on SU(2) invariant ladders in order to answer
the important question which phases are accessible 
for a general choice of tunable couplings within the framework of an effective field 
theory. 
In particular, an incommensurate
phase is postulated for a certain class of ladder systems in the weak coupling limit, 
which has so far received little attention.

\begin{figure}[t!]
\begin{center}
\includegraphics[width=\columnwidth]{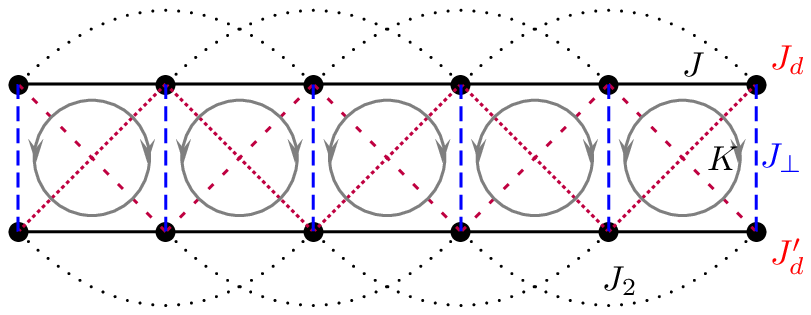}
\caption{Generalized spin ladder with ring exchange interaction $K$, NNN coupling $\jnnn$, 
and two diagonal couplings $\jdga$ and $\jdgb$. }\label{fig-model}
\end{center}
\end{figure}

The generalized SU(2) invariant spin-1/2 ladder is described by the 
Heisenberg Hamiltonian plus a ring exchange interaction 
\BE
H = \sum_{<i,j>} J_{ij} \mathbf{S}_i \cdot \mathbf{S}_j 
+\jrng \sum_{p} (P_{p}+P_{p}^{-1}) ~,    
\label{model}
\EE
where $J_{ij}$ take on values of antiferromagnetic nearest-neighbor coupling $J>0$ 
and next-nearest 
neighbor (NNN) coupling $\jnnn$ in the two chains as well as  
two diagonal couplings $\jdga,~\jdgb$ and a
rung coupling $\jvrt$ between the chains as depicted in Fig.~\ref{fig-model}.
The second term in Eq.~\eqref{model} sums over the 
plaquettes of the system, where $P_{p}$ stands for the cyclic permutation of the 
spins on the four sites of the $p$--th plaquette. 
Such a ring exchange interaction arises from the higher order expansion 
of the Hubbard model in the strongly interacting limit \cite{PhysRevB.37.2353, 
PhysRevB.69.094435} and  
can also be written as a product of two--spin permutations 
$P_p=P_{p_1p_2} P_{p_1p_3} P_{p_1p_4}$, which in terms  of Pauli matrices  
$\bm{\sigma}$ reads, $P_{ij}=(1+\bm{\sigma}_i\cdot\bm{\sigma}_j)/2$ \cite{RevModPhys.55.1}.  

The tunable parameters in Fig.~(\ref{fig-model}) provide a generalization of 
previously studied ladder models.   It has been established that weak 
interchain couplings
give rise to four possible gapped quantum phases, which are characterized by 
correlations in form of singlets \cite{PhysRevB.53.52,PhysRevB.53.8521, PhysRevB.62.14965, JPSJ.64.1967, JPSJ.72.1022,PhysRevB.66.134423, PhysRevB.69.094431, PhysRevB.88.075132,PhysRevB.88.104403}:

\setitemize[0]{leftmargin=10pt,itemindent=10pt}
\begin{itemize}

\item  {\it Rung singlet phase:}  Singlet formation across the rungs for 
an antiferromagnetic rung coupling $\jvrt >0$.
\item  {\it Haldane phase:}  Singlet formation analogous to the so-called AKLT state\cite{Affleck1988} 
in the spin-1 model with a topological string order parameter for $\jvrt <0$.
\item  {\it Columnar dimerized phase:}  Alternating singlet 
formation within each chain on the 
same bonds in both legs for $\jrng <0$.
\item  {\it Staggered dimerized phase:}  Alternating singlet
formation within each chain with a shift of one site between the two legs 
for $\jrng >0$.

\end{itemize}

The rung singlet and Haldane phases have short range correlations, while the 
dimer-dimer correlations in the dimerized phases are long range and break translational 
invariance.
Those four phases with singlet formation are also dominant 
if one considers frustrating couplings $\jdga,\jdgb, \jnnn$, which normally 
enhance quantum fluctuations.  
Such frustrating diagonal coupling have been extensively studied  
in the form of 
a cross coupled ladder (CCL) for $\jdga=\jdgb$ 
\cite{PhysRevB.61.8871,PhysRevB.62.14965,PhysRevLett.93.127202,
PhysRevB.81.064432,PhysRevB.86.075133,PhysRevB.73.224433,PhysRevB.77.205121}
or 
a diagonal ladder (DL) for $\jdga=0$ 
\cite{PhysRevB.59.7973, PhysRevB.72.104435, PhysRevB.77.205121}.  
Both the DL and the CCL can be tuned from a rung singlet phase to a Haldane phase with 
increasing $\jdga/\jvrt$.  Interestingly, the 
columnar dimerized phase also appears 
for the DL for intermediate $\jdga/\jvrt$ \cite{PhysRevB.77.205121}, but such a phase is
debated for the CCL since it would spontaneously break 
translational invariance \cite{PhysRevLett.93.127202,
PhysRevB.81.064432,PhysRevB.77.205121,PhysRevB.86.075133,PhysRevB.73.224433}.
A frustrating in-chain next-nearest neighbor interaction is known 
to cause dimerization in single chains  
for $\jnnn>J_{2c}\approx0.241167 J$
\cite{0305-4470-27-17-012,
PhysRevLett.75.1823, PhysRevB.54.R9612} and therefore 
also stabilizes the corresponding dimer phases in ladder systems 
with next-nearest neighbor coupling
\cite{PhysRevB.73.214427,PhysRevB.84.144407,PhysRevB.89.241104,
PhysRevB.77.214418, 1367-2630-14-6-063019}. 

We now want to answer the question if the enhanced entropy from competing phases may
open the possibility for interesting 
additional phases.  We therefore consider the interplay of all couplings in the full model in
Fig.~(\ref{fig-model}) using 
the framework of a renormalization group (RG) treatment.
As we will see below there are two relevant operators which are responsible for the 
known four phases above, but for $\jdga \neq \jdgb$ the enlarged unit cell allows 
an additional
relevant operator which may dominate in parts of the phase diagram and 
gives rise to an interesting incommensurate phase.

Starting from the established effective field theory description of two decoupled chains, 
the spin operators acquire the following representation in the continuum limit
\cite{affleckleshouches, PhysRevB.46.10866,Ian1986409, tsvelik, book-cft}  
\BE \label{spin operator}
\mathbf{S}(x)\approx
\mathbf{J}(x)+(-1)^{x}~\const~\mathbf{n}(x)~,
\EE
where $\const$ is a non--universal constant \cite{PhysRevB.82.214420}, 
and the lattice constant is set to unity.   The uniform part of the 
spin operator is the sum of the chiral $\mathrm{SU(2)}$ currents of 
the Wess-Zumino-Witten (WZW) model, $\mathbf{J}=\mathbf{J}_{L}+\mathbf{J}_{R}$; 
the staggered part, is related to the fundamental  field $g$ of the WZW model 
via $\mathbf{n} \sim \text{tr} {\bm \sigma} g$, while the trace 
of $g$ gives the dimerization operator, $\epsilon \sim \mathrm{tr}g$, with 
$\epsilon_j = (-1)^j\mathbf{S}_j\cdot\mathbf{S}_{j+1}$.   
Without interchain couplings the field theory can be written for each leg 
separately in the form
\begin{equation}\label{ham-single-chain} 
H_0 =   \frac{2\pi v}{3}\!\! \int\!\! dx \big[ :\!\mathbf{J}_L\cdot \mathbf{J}_L\!\!:
+:\!\mathbf{J}_R\cdot \mathbf{J}_R\!\!:
+\frac{3 \cpla}{2}\mathbf{J}_L\cdot \mathbf{J}_R  
\big]  
\end{equation}
where the first two normal ordered terms are conformally invariant and the last part
represents a backscattering marginal operator.
Without any additional couplings between the two legs of the ladder 
this theory
describes a {\it spin liquid} for each chain 
which is generally unstable to perturbations, however.
The velocity $v\approx \tfrac{\pi J}{2}-1.65 \jnnn$ and the value of the 
marginal coupling $\cpla$ can be tuned by the 
in-chain nearest neighbor coupling $\jnnn$\cite{affleckleshouches,PhysRevB.46.10866,PhysRevB.54.R9612}.


%
We denote the field theories for each chain with an additional index $\eta=1,2$.
The allowed symmetries of the microscopic model dictate that up to 
four additional relevant or marginal operators 
can be generated by the coupling between the legs of the ladder
\BE\label{hpert}
\delta \mathcal{H}= 2 \pi v \int dx \Big(
\cpln \on + \cplc \oc + \cple \oe +\cplb \ob\Big) ~. 
\EE
Here, the 
operator $\on =  \mathbf{n}_1\cdot \mathbf{n}_2$ of scaling dimension 1 describes
the coupling between the antiferromagnetic part of the spins,  
which drives the system into a rung singlet phase for $\cpln>0$ and 
into a Haldane phase for $\cpln<0$.  An effective coupling of the 
dimerization is described by 
the relevant operator $ \oe = \epsilon_1 \epsilon_2$ of scaling dimension 1, which may
drive the system into a staggered ($\cple>0$) or columnar ($\cple<0$) dimerized phase.
Interestingly, for $\jdgb\neq \jdga$ the unit cell is enlarged which allows another 
relevant operator
$ \oc = \mathbf{J}_1\cdot \mathbf{n}_{2}-\mathbf{J}_2\cdot \mathbf{n}_{1}$ 
of scaling dimension 3/2, which is invariant under translation by two sites.
It is not invariant under translation by one site, unless the chain index is also
exchanged, in agreement with the model in Fig.~\ref{fig-model} for $\jdgb\neq \jdga$.
As we will see below this operator can drive the system into 
yet another interesting {incommensurate} phase.
Finally, there is another marginal coupling 
$\ob=\mathbf{J}_{1,L}\cdot\mathbf{J}_{2,R}+\mathbf{J}_{1,R}\cdot\mathbf{J}_{2,L}$ of scaling
dimension 2, which can tip the balance of which relevant operator dominates under
renormalization, which 
in turn determines the phase of the ground state.

The corresponding RG equations of the 
bare coupling constants are determined up to second order 
according to the operator product expansion \cite{book-cardy, book-senechal-2004}  
\BE
\frac{d\lambda_k}{dl} =(2-d_k)\lambda_k-\frac{\pi}{v}\sum_{i,j} C_{ijk} 
\lambda_i \lambda_j~, 
\EE
with $d_k$ the scaling dimension of each operator and the coefficients 
$C_{ijk}$ are obtained by the operator product expansion 
\cite{book-cft, PhysRevB.72.094416}
as discussed in the
Appendix.
For the operator content in Eq.~\eqref{hpert}, 
we arrive at the following RG flow ($\dot\lambda=d\lambda/dl$)  
\begin{subequations}\label{rgFlow}
\BEA
\dot{\cpla} &=& \cpla^2 + \frac{1}{2}\cple^2 -
\frac{1}{2}\cpln^2~, \label{rg-cpla} \\
\dot{\cplb}  &=& \cplb^2 -\cple\cpln
+ \cpln^2~,
\label{rg-cplb} \\
\dot{\cple}  &=& \cple +\frac{3}{2}\cpla\cple
-\frac{3}{2}\cplb\cpln -\frac{3}{2}\cplc^2 ~, \label{rg-cple}\\
\dot{\cpln}   &=& \cpln -\frac{1}{2}\cpla\cpln
-\frac{1}{2}\cplb\cple +\cplb\cpln -\cplc^2 ~, \label{rg-cpln}\\
\dot{\cplc} &=& \frac{1}{2}\cplc -\frac{1}{4}\cpla\cplc
+\cplb\cplc + \frac{1}{2}\cplc\cple -\cplc\cpln \label{rg-cplc}~. 
\EEA
\end{subequations}
The bare coupling constants can be determined from the microscopic model 
using Eq.~(\ref{spin operator}) as also discussed in the
Appendix
\BEA
\cpln & =&  \const^2 \frac{\jvrt-\jdgl}{2\pi v}~, 
\cple = 36\frac{\const^2}{\pi^2}\frac{\jrng}{2\pi v}~,  
\cplc=\const \frac{\jdga-\jdgb}{2\pi v}, \nonumber \\ 
\cpla &=&  1.723(J_{2}-J_{2c}) +\Big(2-\const^2 +\frac{3\const^2}{\pi^2}(1+\const^2) 
\Big)\frac{\jrng}{2\pi v}, \nonumber   
\\\cplb &=&\frac{\jvrt+\jdgl}{2\pi v}
+4\Big(1-\frac{(1+\const^2)^{2}}{\pi^2}\Big)\frac{\jrng}{2\pi v},~ 
\label{bareCouplings}
\EEA
where we have used $\jdgl=\jdga+\jdgb$. 
As expected $\oc$ is forbidden by symmetry for $\jdgb = \jdga$
and stays zero under renormalization in 
Eqs.~\eqref{rgFlow} in this case. 
The system remains approximately scale invariant above an energy scale $\Lambda$ as long
as the coupling constants are small.  
However, as the cutoff $\Lambda(l)=\Lambda_0 e^{-l}$ is lowered
typically one of the coupling constants effectively becomes of order
unity under renormalization, which determines the dominant ground state correlations.
In turn the value of $\Lambda(l)$  at this breakdown point
defines a new intrinsic energy scale, below which scale
invariance is lost and no further renormalization is possible.
Good agreement with 
numerically determined phase transitions has been achieved when using $\const =1$ 
and integrating the RG equations up to a breakdown point where one coupling reaches
$\lambda_*=1$, which is the procedure we use in the following.

\begin{figure}[t!]
\begin{center}
\includegraphics[width=\columnwidth]{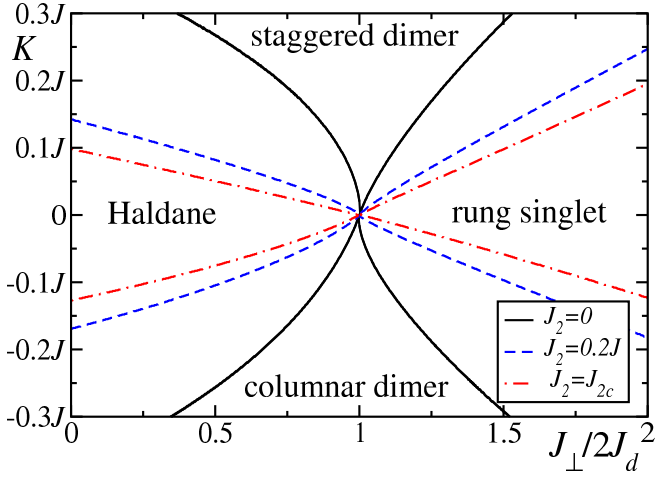}
\caption{Phase diagram of the cross coupled ladder as a function of $\jrng$,  
and the ratio $\jvrt/2\jdga$, for $\jdga=\jdgb=0.2J$, and three values of $J_{2}=0$, $0.2J$, 
and  $J_{2c}$, denoted by  black solid, blue dashed, and  
red dot--dashed lines respectively.} 
\label{pd-CCL}
\end{center}
\end{figure}

The operators with the smaller scaling dimension, i.e.~$\on$ and $\oe$, renormalize 
faster. They will determine the low energy physics in most of the cases in form of
a rung singlet ($\cpln>0$), Haldane ($\cpln<0$), columnar dimer ($\cple<0$), or
staggered dimer phase ($\cple>0$) as already discussed above.  
However, there are regions of the parameter space where the outcome of the renormalization 
procedure is less trivial to predict due to the competition of the frustrating couplings. 
In particular,  if the bare couplings $\cple, \cpln$ are very small or
vanish it is possible that other operators dominate.  
The relevant coupling $\lambda_c$ is of particular interest
in this respect since it may drive the phase into an incommensurate phase as discussed below.
However, 
$\cple$ and $\cpln$ will always be generated by higher order terms in 
the renormalization procedure\cite{PhysRevLett.93.127202} and may still 
dominate the low energy physics depending 
also on the flow of the marginal couplings. 

%

To illustrate the interplay of different coupling constants we first consider the 
cross coupled ladder (CCL) with $\jdga=\jdgb$ in the presence of in-chain frustration $\jnnn$
and ring exchange coupling $K$.   As shown in Fig.~\ref{pd-CCL} the four known phases 
dominate the phase diagram since $\cplc=0$ by symmetry in this case. 
For $\jrng=0$ there is a direct
phase transition from rung singlet to Haldane phase at $\jvrt/2\jdga=1$ with no intermediate 
dimerized phase.  The next nearest 
neighbor coupling $\jnnn$ does not change the basic topology of the phase diagram, 
but has a large effect on the range of the dimer phases.
This is due to the fact that the in-chain frustration $\jnnn$ leads to a reduced starting value 
of $\cpla$ and therefore makes the dimerization operator effectively more relevant.
In fact, it is known that decoupled chains spontaneously dimerize for $\jnnn>J_{2c}$
\cite{0305-4470-27-17-012, PhysRevLett.75.1823, PhysRevB.54.R9612}.

\begin{figure}[t!]
\begin{center}
\includegraphics[width=\columnwidth]{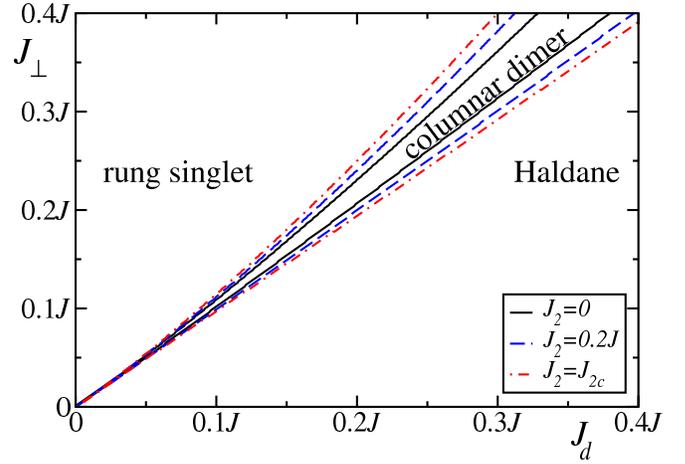}
\caption{Phase diagram of the diagonal ladder  ($\jdgb=0$)
as a function of $\jdga$ and $\jvrt$,  
for $\jrng=0$ and three values of $J_2=0$, $0.2J$, and $J_{2c}$,  denoted by black solid, blue dashed, 
and red dot--dashed lines respectively.}  
\label{pd-DL-JvJd}
\end{center}
\end{figure}

For comparison the phase diagram of the diagonal ladder (DL) with $\jdgb=0$ is shown in 
Fig.~\ref{pd-DL-JvJd} for $\jrng=0$
and three values of the NNN interaction, $J_2=0, 0.2J, J_{2c}$
as function of the two interchain couplings $\jvrt$ and $\jdga$. 
The symmetry properties of this model are fundamentally different from the CCL, because the 
larger unit cell permits the presence of the operator $\oc$, which 
plays an important role in generating the
dimer operator $\oe$ 
under the renormalization in Eq.~(\ref{rgFlow}). 
Interestingly, in contrast to the CCL, therefore an 
intermediate columnar dimerized phase exists even {\it without} ring exchange, where the
generated operator $\oe$ dominates the correlations. 
This phase again becomes larger in the presence of in-chain frustration $\jnnn$ for reasons
mentioned above.

\begin{figure}[t!]
\begin{center}
\includegraphics[width=\columnwidth]{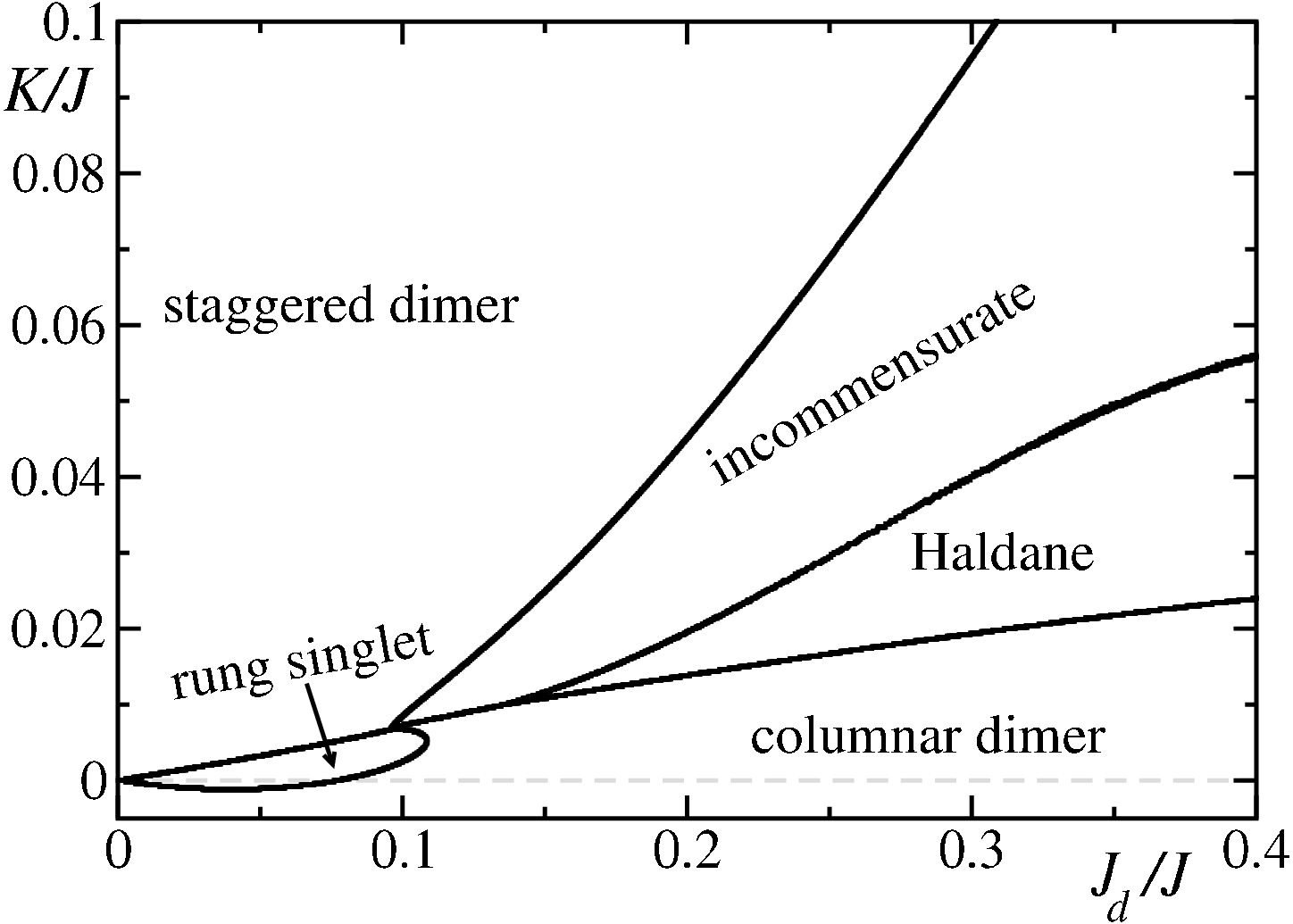}
\caption{Phase diagram of the diagonal ladder ($\jdgb=0$) as a function of $\jdga$ and $\jrng$,  
for $\jvrt/\jdga=1.1$ and $\jnnn=0.2J$.}   
\label{pd-DL-JdJr}
\end{center}
\end{figure}

One interesting aspect of the phase diagram in Fig.~\ref{pd-DL-JvJd} is the fact that 
there is no phase which is generated directly by the relevant operator $\oc$ even in cases
where its initial bare coupling constant $\cplc$ may be largest.  This invites interesting questions: 
First of all, is it ever possible to generate phases which are characterized by a dominant 
coupling $\cplc$ in the extended parameter space?  
And secondly, if such a phase exist, what dominant correlation 
would be expected?

The first questions is answered by finding suitable parameters for such a phase by taking 
advantage of canceling out the effect from competing dimer phases.
We understand that the dimerized phase in Fig.~\ref{pd-DL-JvJd} is created by $\oe$, 
which grows beyond bounds after it  
is generated in second order by the initial bare coupling $\cplc$. To suppress this
generation it is possible to add a small ring exchange $K$, which initially pushes
the value of $\cple$ to be slightly positive in Eq.~(\ref{bareCouplings}).  The analogous reasoning applies for
$\cpln$, where a small positive bare value can be achieved by 
choosing $\jvrt$ slightly larger than $\jdga$. 
Accordingly, we show the phase diagram for the diagonal ladder with $\jvrt = 1.1 \jdga$
and $\jnnn=0.2J$ as a function of $K$ and $\jdga$.  In this case there is indeed an extended
region with this additional phase for $K>0$, which we describe as {\it incommensurate}
for reasons that will be explained below.  This incommensurate phase separates 
the staggered dimer from the columnar dimer phase.

\begin{figure}[t!]
\begin{center}
\includegraphics[width=\columnwidth]{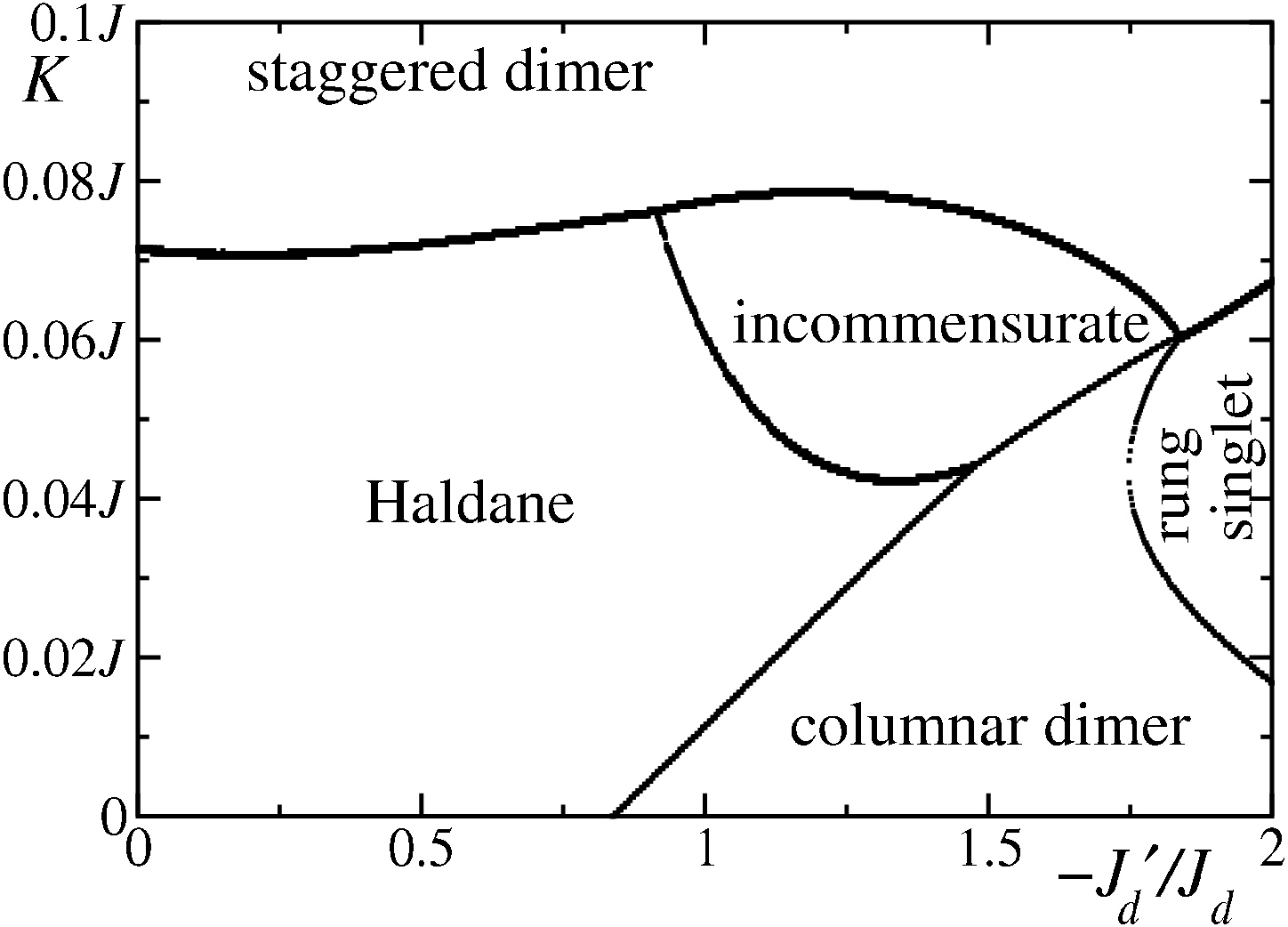}
\caption{Phase diagram as a function 
of $-\jdgb/\jdga$ and $\jrng$,   
for $\jvrt=0$, $\jnnn=J_{2c}$ and $\jdga=0.2J$.}  
\label{pd-DDL}
\end{center}
\end{figure}

To illustrate another example of possible parameters for this phase, we consider
a general ratio of the diagonal couplings $\jdgb/\jdga$ in Fig.~\ref{pd-DDL} for
$\jvrt=0$, $\jnnn=J_{2c}$  and $\jdga=0.2J$.  
According to Eq.~(\ref{bareCouplings}) the initial value of $\cplc$ 
becomes strongest for $\jdgb=-\jdga$, while the other initial couplings are zero for
$\jvrt=\jrng=0$ and $\jnnn=J_{2c}$. Indeed  the intermediate incommensurate phase appears 
for small positive $\jrng$ and $\jdgb \approx -\jdga$, again separating the staggered and
columnar dimer phases.

To answer the question about the nature of the phase for strong 
$\oc = \mathbf{J}_1\cdot \mathbf{n}_{2}-\mathbf{J}_2\cdot \mathbf{n}_{1}$  it 
is possible to invoke a self-consistency argument, analogous to a chain mean field theory\cite{0305-4470-27-22-009,PhysRevB.11.2042,PhysRevLett.89.047202}.
Clearly, the operator $\oc$ causes an alternating magnetic order $\mathbf{n}$ in one chain
to induce a collinear ferromagnetic order $\mathbf{J}$ in the other chain and vice versa. 
Assuming small finite order parameters we can write for the expectation values
\BEA 
\langle \mathbf{J}_2 \rangle & = & -\cplc \chi_0 \langle \mathbf{n}_1 \rangle; \ \ 
\langle \mathbf{n}_1 \rangle = -\cplc \chi_1 \langle \mathbf{J}_2 \rangle \nonumber \\
\langle \mathbf{J}_1 \rangle & = & \cplc \chi_0 \langle \mathbf{n}_2 \rangle; \ \ 
\langle \mathbf{n}_2 \rangle = \cplc \chi_1 \langle \mathbf{J}_1 \rangle \label{cmft}
\EEA
where $\chi_0$ is the uniform susceptibility and $\chi_1$ is the alternating susceptibility.
While $\chi_0$ is finite \cite{PhysRevLett.73.332}, $\chi_1$ diverges with $1/T$ \cite{PhysRevB.11.2042,PhysRevLett.89.047202}, so 
there will always be a low enough energy scale at which 
$\lambda_c^2 \chi_0 \chi_1 =1$ results in a self-consistent order.   Clearly, at this point 
both the alternating antiferromagnetic order 
and the collinear ferromagnetic order become finite in both chains. Since it is known
that a small collinear ferromagnetic part effectively shifts the wavelength of the
antiferromagnetic order\cite{PhysRevB.60.1038,PhysRevLett.75.934}, we conclude that this 
behavior corresponds to 
an incommensurate state.  
Semi-classically $\mathbf{n}_{1(2)}$ is perpendicular to $\mathbf{J}_{1(2)}$, so therefore
also $\mathbf{n}_{1}$ is perpendicular to $\mathbf{n}_{2}$ in this mean-field argument.

Incommensurate behavior due to frustration has been much discussed in the literature, 
not only for strongly frustrated chains \cite{PhysRevB.54.9862} and coupled 
ladder systems \cite{PhysRevB.84.144407,PhysRevLett.81.910,0953-8984-10-5-017}, 
but more recently also for 
anisotropic triangular lattices\cite{0295-5075-87-3-36002, PhysRevB.84.245130, PhysRevB.86.184421, PhysRevLett.117.193201}.
However, the incommensurate phase we have predicted in this paper arises for {\it small}
frustration and interchain couplings.  It is generated by a complex combination of competing 
instabilities, which require only a small perturbation from the spin-liquid 
fixed point of decoupled chains.  This opens the door for the search also in 2D
systems for corresponding incommensurate phases near instabilities.

In conclusion, we have examined a general SU(2) invariant ladder model 
with focus on frustration and ring exchange.  For small coupling strengths
the renormalization flow to four known phases can be quantitatively examined.  In addition, 
we predict that there is an instability to an incommensurate phase in parts of the 
parameter space.  In contrast to other models with incommensurate behavior, this phase
can be observed even for very small values of frustration and inter-chain couplings.
While frustration is an important ingredient for enhanced quantum fluctuations in order
to generate the phase, the 
underlying instability towards incommensurate order is already present in the field theory
of any weakly coupled chains with a broken translational symmetry $\jdga\neq \jdgb$.  
It therefore appears that
an essential aspect for the observed incommensurate behavior is the  
enlarged unit cell, which in turn provides an important clue what choice of couplings may
be promising to show corresponding phases in a 2D generalization of the model.

\begin{acknowledgments}
A.~M.~ acknowledges fruitful discussions with N.~P.~Konstantinidis.  This work received 
financial support from the Deutsche Forschungs Gemeinschaft (DFG) through the
SFB-TR 49, the SFB-TR 173, and the SFB-TR 185.
\end{acknowledgments}
%
%
%
%
\section{Appendix}
 
In this appendix additional information for the derivation of the renormalization 
group equations is given and the bosonization of the four spin interaction is discussed.

\subsection{Field theory and renormalization group flow}\label{app-ope}
For the derivation of the bosonization formulas and the operator product expansion (OPE) 
it is useful to consider an interacting spinful Fermion model as the underlying physical 
realization where only the spin channel will be considered in the low energy limit.  
For the half-filled Hubbard model the charge channel is gapped and the 
Heisenberg couplings considered in the paper 
corresponds to the spin channel.

The spin currents are therefore conveniently expressed in terms of left- and right-moving 
Fermion operators
\BE
J_\kappa^a (z_\kappa)  
= :\psi_{\kappa \eta}^{\dag}\frac{\sigma_{\eta \eta'}^a}{2}\psi_{\kappa \eta'}:(z_\kappa) ~,
\EE
where $\sigma^a$ are the Pauli matrices with a summed over spin-index 
$\eta=\uparrow,\downarrow$, and $\kappa=R,L$ denotes the chirality. 
 The chiral complex coordinates are $z_{L/R}=\pm ix + v\tau$. 
The dimerization and staggered magnetization operators are given by \cite{PhysRevB.72.094416}
\BEA
\epsilon(z) &\sim & \frac{i}{2}\big[
:\psi_{R\eta}^{\dag} \psi_{L\eta}:(z)-:\psi_{L\eta}^{\dag} \psi_{R\eta}:(z) \big]~,\\
n^a (z) &\sim & \frac{1}{2}\sigma_{\eta\eta'}^a\big[ :\psi_{R\eta}^{\dag}\psi_{L\eta'}:(z)+ 
:\psi_{L\eta}^{\dag}\psi_{R\eta'}:(z)\big]~,   \nonumber 
\EEA 
where $z$ implies a dependence on both chiral variables $z_R$, $z_L$ in this case.  
\par  
The OPEs between $J_\kappa^a$, $\epsilon$,  and $n^b$ can be calculated using 
Wick's theorem \cite{book-cft} and the  two point correlation function 
\BE 
\langle \langle \psi_{\kappa \eta}(z_\kappa)\psi_{\kappa'\eta'}^\dag(w_{\kappa'})\rangle \rangle =
\delta_{\kappa\kappa'}\delta_{\eta\eta'}\frac{\gamma}{z_\kappa-w_\kappa}~, 
\EE
where we choose a normalization of $\gamma=1/2\pi$.  
Keeping all relevant terms, the important  OPEs are 
 \cite{PhysRevB.72.094416}  
\BEA\label{fundamentalopes}
J^a_\kappa(z_\kappa)J^b_{\kappa'}(w_{\kappa'})&=&\delta_{\kappa\kappa'}\left[ 
\frac{(\gamma^2/2)\delta_{ab}}{(z_\kappa-w_\kappa)^2} +i\epsilon_{abc}\gamma 
\frac{J_\kappa^c(w_\kappa)}{z_\kappa-w_\kappa}\right] \nonumber \\
J^a_\kappa(z_\kappa) \epsilon(w) & =&  i\kappa \frac{\gamma/2}{z_\kappa-w_\kappa} 
n^a(w) \nonumber \\ 
J^a_\kappa(z_\kappa) n^b(w) &=& i\frac{\gamma/2}{z_\kappa- w_\kappa} 
[\epsilon_{abc} n^c(w) -\kappa\delta_{ab} \epsilon(w)] \nonumber \\
\epsilon(z)\epsilon(w)&=& \frac{\gamma^2}{|z-w|}-|z-w| \mathbf{J}_R\cdot \mathbf{J}_L(w)  \\ 
n^a(z)\epsilon(w)&=&-i\gamma |z-w| \left[ 
\frac{J_R^a(w_R)}{z_L- w_L}-\frac{J_L^a(w_L)}{z_R-w_R}
\right] \nonumber \\ 
n^a(z)n^b(w)&=&|z-w| \Big[\frac{\gamma^2 \delta_{ab}}{|z-w|^2} 
+i\epsilon_{abc} \gamma \nonumber \\  &\times &  \left[
\frac{J_R^c(w_R)}{z_L- w_L}+\frac{J_L^c(w_L)}{z_R-w_R}
\right] +\hat Q^{ab}(w)\Big] \nonumber
\EEA 
where, $\delta_{ab}$ is the Kronecker $\delta$--function, and $\epsilon_{abc}$ the Levi Civita symbol. Here 
\BE
\hat Q^{ab} = \frac{1}{2} \sigma_{\eta\eta'}^a \sigma_{\tau\tau'}^b
\psi_{R\eta}^\dag \psi_{L\eta'} \psi_{L\tau}^\dag \psi_{R\tau'} ~ ,
\EE
denotes the zeroth order 
contraction between the fermionic fields. 
After freezing out the gapped charge degrees of freedom 
only the trace of this operator is relevant for the calculation of the 
renormalization group flow,  which reads
\BE
\hat Q^{aa} = \mathbf{J}_R \cdot \mathbf{J}_L  ~. 
\EE
\par 
The evolution of the bare couplings is determined 
from the OPEs of the perturbing operators using   \cite{book-cardy} 
\begin{equation}\label{couplingevolution}
\frac{d\lambda_k}{dl} =(2-d_k)\lambda_k-\frac{\pi}{v}\sum_{i,j} C_{ijk} \lambda_i \lambda_j~, \quad
\end{equation}   
where $d_k$ is the scaling dimension of the operator, $v$ is the velocity, and $C_{ijk}$ 
is the coefficient extracted from the OPE $O_i(z)O_j(w)\sim O_k(w)$.    
Using Eqs. \eqref{fundamentalopes} and \eqref{couplingevolution}  we arrive 
at the renormalization group equations in the main text.  
\subsection{Four spin interactions}
\par 
In this part we present the bosonization of the four--spin interactions. 
The ring exchange interaction is given by 
\BE\label{ringExchange}
H= \jrng \sum_p \left(h_p+\frac{1}{4}\right)~, 
\EE
where $p$ sums over the plaquettes of the system and the local energy operators are given by  
\BEA 
h_p&=&\mathbf{S}_{p_1} \cdot\mathbf{S}_{p_2}+\mathbf{S}_{p_3} \cdot\mathbf{S}_{p_4}  
+ \mathbf{S}_{p_1} \cdot\mathbf{S}_{p_4}+\mathbf{S}_{p_2} \cdot\mathbf{S}_{p_3}  \nonumber \\
&+& \mathbf{S}_{p_1} \cdot\mathbf{S}_{p_3}+\mathbf{S}_{p_2} \cdot\mathbf{S}_{p_4} 
+ 4(\mathbf{S}_{p_1} \cdot\mathbf{S}_{p_2})(\mathbf{S}_{p_3} \cdot\mathbf{S}_{p_4}) \nonumber \\
&+& 4(\mathbf{S}_{p_1} \cdot\mathbf{S}_{p_4})(\mathbf{S}_{p_2} \cdot\mathbf{S}_{p_3}) 
- 4(\mathbf{S}_{p_1} \cdot\mathbf{S}_{p_3})(\mathbf{S}_{p_2} \cdot\mathbf{S}_{p_4}) ~.  \nonumber
\EEA
The indices  $p_1-p_4$ count clockwise the spins of the $p$-th plaquette. 
The Hamiltonian consists of products of spin operators, $H_2$, and 
products of  pairs of spin operators, $H_4$. 
\par 
The    four--spin interactions are  either on the same leg of the ladder or on the 
rungs of the ladder. The leg component $H_{L}$ and the rung $H_{R}$ are 
given by 
\BEA
H_{L} &=& 4\jrng \sum_{j}  (\mathbf{S}_{1,j}\cdot \mathbf{S}_{1,j+1}) 
(\mathbf{S}_{2,j}\cdot\mathbf{S}_{2,j+1})~, \\
H_{R} &=& 4\jrng \sum_{j}  (\mathbf{S}_{1,j}\cdot \mathbf{S}_{2,j})
(\mathbf{S}_{1,j+1}\cdot\mathbf{S}_{2,j+1}) \\ 
&-& 4\jrng \sum_{j} (\mathbf{S}_{1,j}\cdot \mathbf{S}_{2,j+1})
(\mathbf{S}_{2,j}\cdot\mathbf{S}_{1,j+1}) ~. \nonumber 
\EEA
Spin operators are substituted in the continuum with, 
\BE
\mathbf{S}_{j}/a \rightarrow \mathbf{S}(x)\approx\mathbf{J}(x)+(-1)^{x}\const \mathbf{n}(x)~, 
\EE
and the product of two spin operators on the same chain becomes 
with the help of Eq.~\eqref{fundamentalopes}, 
\BEA \label{HDDRR1}
aS^a(x) S^b(x+a)  &\approx&  - (a\const)^2  \hat Q^{ab}(x) 
+(-1)^x  2\gamma \const  \delta_{ab}  \epsilon(x)   \nonumber \\ 
+~\gamma(1+a\const^{2}) \hspace{-10pt}&& \hspace{-10pt} \left[
\epsilon_{abc} \left[J_{R}^c(x)-J_{L}^c(x)\right]
-\frac{\gamma}{a} \delta_{ab}
\right]~.   
\EEA
Using this equation for the products of spin operators in the leg part 
we arrive at 
\BEA \label{legFourSpin}
H_{L}&\approx& 12a \jrng \gamma^{2} \const^{2} \nonumber  \\
 &\times&\int dx\left[\left(a+(a\const)^{2}\right)\oa(x) + 12  \oe(x) \right]~. 
\EEA 
The rung part of the four spin interactions can be written as 
\BEA 
H_{R} &\approx& 4 a^3 \jrng \int dx S_{1}^a(x)S_{1}^b(x+a)  \\ 
&\times& [S_{2}^a(x)S_{2}^b(x+a)-S_{2}^b(x)S_{2}^a(x+a)] \nonumber ~, 
\EEA
where due to the relative minus sign and the $\delta-$functions, relevant terms are 
eliminated,  and only interchain marginal terms survive 
\BE \label{rungFourSpin}
H_{R} \approx -16 a \jrng  \gamma^2 \left(1+a\const^{2}\right)^{2} 
\int dx \ob(x) ~. 
\EE
Combining Eqs.~\eqref{legFourSpin} and \eqref{rungFourSpin} and carrying out a 
trivial calculation for the two spin interactions $H_2$,
\BE
H_2\approx a\jrng \int dx \left[(2-a\const^2)\oa(x)+4\ob(x)\right]~, 
\EE
one arrives at the bosonized ring exchange interaction in the continuous limit 
\BEA\label{hamRingExchange}
H &\approx& a \jrng \int dx \Bigg[\left(2-a\const^2 +\frac{3a\const^{2}}{\pi^2}(1+ a\const^2) \right) \oa\nonumber \\
&+& 4\left(1-\frac{(1+a\const^2)^{2}}{\pi^2}\right)\ob +36\frac{\const^2}{\pi^2}\oe\Bigg]~.   
\EEA  
which is used to determine the corresponding bare coupling strengths in the main text.
%
%
%
%
%
%
%
%
%

%
%
%
\end{document}